\documentclass[useAMS,usenatbib]{mn2e}
\UseRawInputEncoding 
\usepackage{graphicx,epsfig,amssymb,amsmath,layout,verbatim,rotating,calc,mathrs
fs,natbib}
\usepackage{graphics}
\usepackage{rotate}
\usepackage{float} 
\usepackage{txfonts}
\usepackage{pifont}
\usepackage{bbding}
\usepackage{xcolor, soul}
\usepackage{ulem}
\sethlcolor{yellow}
\def\O{$\Omega$}
\input wasyfont

\title[The phase space around $L_{1,2}$]{The Phase Space Structure in the vicinity of 
vertical Lyapunov orbits around $L_{1,2}$ in a barred galaxy model}
\author[M. Katsanikas \& P.A. Patsis]
{M.~Katsanikas,$^1$\thanks{mkatsan@academyofathens.gr (MK)}
P.A.~Patsis,$^{1}$\thanks{patsis@academyofathens.gr (PAP)}\\
$^1$Research Center for Astronomy, Academy of Athens, Soranou Efessiou 4, GR-115
27, Athens, Greece\\
}
\date{Accepted ..........Received .............;in original form ..........}

\begin{document}
\maketitle

\label{firstpage}
 
\begin{abstract}
We study the phase space structure and the orbital diffusion from the vicinity 
of the vertical Lyapunov periodic orbits around the unstable Lagrangian points 
$L_{1,2}$ in a 3D barred galaxy model. By perturbing the initial conditions of 
these  periodic orbits, we detected the following five types of orbital 
structures in the 4D spaces of section: (i) Ring-like structures, sticky for 
large time intervals to the unstable invariant manifolds of the simple and 
double unstable vertical Lyapunov periodic orbits. (ii) 2D  tori
belonging to quasi-periodic orbits around stable periodic orbits 
existing in the region. They are associated either with vertical stable periodic orbits 
around $L_{4, 5}$ or with ``stable anomalous'' periodic orbits. (iii) Orbits 
sticky for large time intervals to these tori, forming ``sticky tori'', before 
they slowly depart from them. (iv) Clouds of points that have a strong chaotic 
behavior. Such clouds of  consequents have slow diffusion speeds, because they 
are hindered by the presence of the tori around the ``stable anomalous'' 
periodic orbits. (v) Toroidal zones consisting of points that stick for long 
time on the unstable invariant manifolds of the ``unstable anomalous'' periodic 
orbits. By continuing the integration, we find that eventually they become strongly 
chaotic, retaining however small diffusion speeds, due to the presence of the 
tori around the stable anomalous periodic orbits.

\end{abstract}


\section{Introduction}
\label{sec:intro}

In this paper we study the phase space in the vicinity of ``vertical'' 
Lyapunov periodic orbits around the unstable Lagrangian points 
$L_{1,2}$ of a three dimensional (3D) model for barred galaxies. 

There are two families of (Lyapunov) periodic orbits (POs) around  $L_{1,2}$, 
namely the planar (pLPO) \citep[see e.g.][]{rm09,tkec09,srm16} and the vertical 
(vLPO) family \citep[][]{op98, rm09}. The latter orbits are called ``vertical'', 
due to their orientation with respect to the equatorial plane. In effect, there 
are two vLPO families, one around $L_1$ and another around $L_2$, thus whenever 
needed to be distinguished, we will call them vLPO1 and vLPO2, respectively. 
The same occurs for the  equilibrium points $L_{4,5}$ and we will call the corresponding
vLPO families as  vLPO4 and vLPO5, respectively.  

We want to examine the phase space close to vLPOs, at the ends of a bar. 
In order to achieve this goal, we investigate the role of stickiness in three 
dimensions around equilibrium points, as well as the role of the phase space 
environment in hindering orbits with initial conditions in the neighborhood of 
the vLPO's to diffuse to large chaotic seas. Thus, knowledge of the presence and 
stability of other  POs in the region is important in our study. 
Furthermore, we want to study the relationship between diffusion and chaoticity of the orbits, 
which is a general problem in dynamical astronomy and nonlinear dynamics.

A standard method for studying the structure of phase space is by means of 
spaces of section. As described in standard textbooks about nonlinear phenomena 
in Hamiltonian systems \citep[see e.g.][section 2.11.11]{gcobook}, a 3D galactic Hamiltonian 
model has a 6D phase space, which is reduced by the Jacobi constant in 
time-independent models to a 5D space. By considering a space of section, we 
establish a well defined 4D symplectic map (Poincar\'e map) \citep{poinc92} in a 
4D phase space. Eventually, the study of the 6D phase space of a 3D autonomous 
Hamiltonian system ends up to the study of the 4D phase space of a 4D symplectic 
map. 

The basic problem of the study of the 4D phase space is its visualization. Several methods have been proposed in
 the past for the visualization of the 4D phase space: 2D and 3D projections
 \citep{Contopoulos89,vetal97}, stereoscopic
 projections \citep{Froeschle70,Contopoulos82}, 2D and 3D slices of 3D
 subspaces \citep{Froeschle70,Froeschle72,retal14}, the method of color and rotation \citep{pz94,kp11}  
 and the method of Lagrangian Descriptors \citep{Agaoglou19,Agaoglou20,Agaoglou20b}. As in 
many of our previous studies, in our present work we use the method of color and 
rotation. Briefly, in this 
method we plot an orbit in a 3D subspace of the 4D phase space of the 4D 
Poincar\'e map. Then, we use standard graphical packages to rotate the 3D figure 
and understand its shape. Finally, we color every point of this 3D projection, 
i.e. each consequent,
according to its fourth coordinate. The rotation helps us in order to understand 
the 3D topology of the structures in the phase space. The color variation indicates if 
this topology  exists in the 4D space or not. This method helps us to 
distinguish between  order and chaos by relating specific 3D structures 
accompanied by   
specific patterns of color, to specific cases of orbital stability (see section 
\ref{sec:po}). The method has been used in galactic Dynamics 
\citep{kp11,kpp11,kpc11,kpc13,pk14a,pk14b}, in a 4D symplectic map 
\citep{zkp13}, in Astrodynamics \citep{ge13} and in a system of spinning test 
particle moving in the spacetime background of a Kerr black hole \citep{lg16}.  

An important phenomenon that is crucial for the study of the phase space, and plays 
crucial role in our paper, is the phenomenon of stickiness. There are two kinds of stickiness:

\begin{enumerate}
 \item {\bf Stickiness on  tori:} In the neighborhood of a stable periodic orbit, the KAM (Kolmogorov-Arnold-Moser) theorem \citep[see][]{k54,a63,m62}, guarantees the existence of $N$-dimensional invariant tori in Hamiltonian systems with $N$ degrees of freedom. When the perturbation increases a particular torus is destroyed and it becomes a cantorus, a Cantor set of points that is nowhere dense, \citep[see for example][section 2.7]{gcobook}. This object has a countable infinity of gaps. When a last KAM torus is destroyed, the small chaotic regions that were inside it  are connected with the large chaotic sea that surrounds this last KAM torus. This can happen after a long integration time. This means that the region just inside a cantorus is characterized by a large density of points for long time intervals before these points start to diffuse in the large chaotic sea. This is the phenomenon of stickiness and the formed region is called a sticky torus. Stickiness has been extensively studied in Hamiltonian systems of two degrees of freedom \citep[see for example][]{gcobook} and in Hamiltonian systems of three degrees of freedom \citep[see for example][]{kp11}. 
 
 \item {\bf Stickiness in chaos:} In the neighborhood of unstable periodic orbits, there are two kinds of invariant manifolds, the unstable and stable invariant manifolds \citep[see for example][section 2.5]{gcobook}. All orbits with initial conditions on these objects approach the periodic orbit (stable manifold) and move away from the periodic orbit (unstable manifold). In some cases, orbits are stuck for a long time interval in a specific chaotic region of the phase space far away from the invariant tori. This is due to the fact that these orbits are stuck for a long time interval on the unstable invariant manifolds of unstable periodic orbits \citep[see][]{ch13}. This is the phenomenon of stickiness in chaos and it has been studied in Hamiltonian systems with two \citep[][]{ch13} and three  \citep[][]{kpc13} degrees of freedom.    
\end{enumerate}

The phenomenon of stickiness refers to chaotic orbits that remain in a specific region of the phase space for a significant time interval. It has to be distinguished from another category of chaotic orbits, that remain in a specific region of the phase space,  which is the partially chaotic orbits that obey one isolating integral besides the energy and are bounded by regular orbits \citep[see][]{muz17,muz18,ca20}. Latter, it will not be discussed in the present paper.

In section \ref{sec:model}, and in section \ref{sec:po} we briefly present the 
model which we use and some definitions related to the stability and the 
morphology of periodic orbits. In section \ref{sec:start}, we study the phase 
space of our system using the method of 4D spaces of section and we compare our 
results with  Lyapunov Characteristic Numbers. In section \ref{sec:start1}, 
we compute the diffusion speed for the different types of phase space objects we encountered 
in our study. Finally, we discuss our results and we present our conclusions in  section \ref{sec:conclusions}.

\section{The model}
\label{sec:model}
For the purposes of the present paper, we will continue using the Ferrers bar 
model we have already used in several papers in the past \citep[e.g.][]{spa02a, 
pk14a, path19, p22}. It is a popular and extensively studied model for 3D bars, 
initially used by \cite{pf84}. The general model consists of a Miyamoto disc, a 
Plummer bulge and a 3D Ferrers bar. As this part of information is repeated in 
all the above mentioned papers, we avoid giving it again here and we restrict 
ourselves in a general description of the model and in giving the values of the 
parameters we used. For details the reader is referred to the above mentioned papers.

The Miyamoto disc has horizontal and vertical scale lengths, $A$ and $B$ 
respectively, while its total mass is $M_{D}$. The total mass of the 
spheroidal Plummer bulge is $M_{S}$ and its scale length is $\epsilon_{s}$.
The numerical values of these parameters are given in in Table~\ref{tab:models}.


%



For the axes of the Ferrers bar we set $a:b:c = 6:1.5:0.6$, as in \cite{pf84}, 
while is mass is \( M_{B} \). 
The masses of the three components satisfy \( G(M_{D}+M_{S}+M_{B})=1 \), where 
$G$ is the gravitational constant. 
The length unit has been taken as 1~kpc, the 
time unit as 1~Myr and the mass unit as $ 2\times 10^{11} M_{\odot}$. If not 
otherwise indicated, the lengths mentioned everywhere in the paper are in kpc. 
Units will be not repeated on the axes of the figures. 

The total potential of our model is given in Cartesian coordinates $(x,y,z)$ by 
\begin{equation}
\Phi(x,y,z)= \Phi_D + \Phi_S + \Phi_B, 
\end{equation}
where $\Phi_D, \Phi_S$ and $\Phi_B$ are the potentials of the disk, the bulge (spheroid)
and the bar respectively. 

The 3D bar is rotating around its short $z$ axis counterclockwise, with an  angular speed $\Omega_{b}$. The x axis is the
intermediate and the y axis the long one. The equations of motion are derived 
from the Hamiltonian governing the motion of a  test-particle. It can be written in the form:

\begin{equation}
H= \frac{1}{2}(p_{x}^{2} + p_{y}^{2} + p_{z}^{2}) +
    \Phi(x,y,z) - \Omega_{b}(x p_{y} - y p_{x}),
\end{equation}
where $p_{x},~ p_{y},$ and $p_{z}$ are the canonically conjugate momenta. We
will hereafter denote the numerical value of the Hamiltonian by $E_j$ and
refer to it as the Jacobi constant or, more loosely, as the `energy'.


The parameters referring to the masses of the three components, the pattern 
speed, the $E_j$'s of the important for our study resonances and the corotation 
distance, can also be found in Table~\ref{tab:models}.


\begin{table*}
\caption[]{The parameters used in our model: $G$ is the gravitational constant, 
$M_D$, $M_B$, $M_S$ are the masses of the disc, the bar and the bulge 
respectively, $A$ and $B$ the horizontal and vertical scale lengths of the 
Miyamoto disc respectively, $\epsilon_s$ is the scale length of the spheroidal 
bulge, \O$_{b}$ is the pattern speed of the bar,  $E_j(L_{1,2})$, $E_j(L_{4,5})$ are
the values of energy of  the equilibrium points $L_{1,2}$ and  $L_{4,5}$ and $R_c$ is the 
corotation radius}
\label{tab:models}
\begin{center}
\begin{tabular}{ccccccccccc}
$GM_D$ & $GM_B$ & $GM_S$ & $A$ & $B$ & $\epsilon_s$ & $\Omega_b$ & 
$E_j(L_{1,2})$ & $E_j(L_{4,5})$ & $R_c$ \\ 
\hline
  0.72 &  0.2  & 0.08 & 3 & 1 & 0.4 &  0.054 & -0.2029 & -0.1976& 6.31 \\

\hline
\end{tabular}
\end{center}
\end{table*}

\begin{table*}
\caption[]{The initial conditions of the periodic orbits  of the families   v$LPO1,2$ , v$LPO4,5$, $z$-axis family, sao and uao in the Poincar\'e section $z=0$ with $p_z>0$ for $E_j=-0.165$.  }
\label{tab:models1}
\begin{center}
\begin{tabular}{ccccc}
 Family  & $x$ & $y$ & $p_x$ & $p_y$ \\
    \hline
    v$LPO1$   & $0$ & $6.03566990 $ & $-0.2549905669$ & $0$ \\
     v$LPO2$  & $0$ & $-6.03566990$ & $0.2549905669$ & $0$ \\
     				  
	$z$-axis family & $0$ & $0 $ & $0$ & $0$ \\
\hline
     v$LPO4$   & $5.7038012135 $ & $0$ & $0$ & $0.2551977174$ \\
     v$LPO5$   & $-5.7038012135 $ & $0$ & $0$ & $-0.2551977174$ \\
\hline
 sao and its symmetric   & $2.1463697858  $ & $0$ & $0$ & $ -0.2468369233$ \\
                         & $-2.1463697858  $ & $0$ & $0$ & $ 0.2468369233$ \\
\hline
 uao and its symmetric   & $0 $ & $2.1265112564$ & $ 0.4100251830$ & $0$ \\
                         & $0 $ & $ -2.1265112564$ & $ -0.4100251830$ & $ 0$ \\
\hline
\end{tabular}
\end{center}
\end{table*}


%


\section{Periodic orbits and their stability}
\label{sec:po}

The space of section of 3D systems is 4D. In order to study the vertical 
periodic orbits around the Lagrangian points $L_{1,2}$ we use the space of 
section $z$=0, $p_{z} > 0$ and we integrate the equations of motion for a 
given value of the Hamiltonian. Our initial conditions are 
$(x_{0},p_{x_0},y_{0},p_{y_0})$. We find the initial conditions for a PO by 
using a Newton iterative method. We consider a PO as been located, when the 
initial and final coordinates coincide with an accuracy of at least 10$^{-11}$. 
We used a fourth order Runge-Kutta scheme for the numerical integration of the periodic 
orbits, keeping the relative error in the energy  smaller than 10$^{-15}$.

In order to compute the stability of a PO we follow the method of 
Broucke (1969). For this,  we first consider small deviations from its initial 
conditions in the space of section  $z$=0,  $p_{z} > 0$ and we integrate 
this ``perturbed'' orbit to the next upward intersection with the space of 
section. In this way a 4D map $T: \mathbb{R}^{4} \to \mathbb{R}^{4}$ is 
established, which relates the initial with the final point. The relation of the 
final deviations of this, neighboring to the periodic orbit,  to the 
initially introduced deviations, can be written in vector form as: 
$\vec{\xi}=M\,\vec{\xi_{0}}$.  Here $\vec{\xi}$ is the final deviation, 
$\vec{\xi_{0}}$ is the initial deviation and $M$ is a $4 \times 4$ matrix, 
called the monodromy matrix. The eigenvalues of the monodromy matrix can determine the 
kind of linear stability of a periodic orbit. If a periodic orbit has two pairs of eigenvalues 
on the unit circle,  it is  stable. Otherwise, the periodic orbit is  unstable. 
There are three kinds of instability for periodic orbits: simple instability, double instability, and complex instability. Simple and double instability are the cases where the periodic orbits 
have one pair of eigenvalues on the unit circle and one on the real axis (simple instability) or 
two pairs on the real axis (double instability),  as described in \citet{cm85}. Finally, complex instability  
\citep[see eg.][]{cm85, pf85a, pf85b, z93, pz90, sb21, p22,jo04} is the case in which the eigenvalues of 
the periodic orbits form a  complex quadruplet that is off the unit circle.


In order to have an overall impression for the dynamical behaviour of the two 
vLPO families, which have as origin either of the Lagrangian points $L_{1,2}$, 
we computed their orbits and we followed the evolution of their stability by 
means of the Broucke method. These orbits have on the 4D space of section initial 
conditions $(x_{0},p_{x_0},y_{0},p_{y_0})$ with   $(0,y>0,p_x<0,0)$ and 
$(0,y<0,p_x>0,0)$ respectively. Following the evolution of their $b_1$ and $b_2$ 
indices \citep{cm85}, we found that they are in general simple unstable, with a rather small 
double unstable energy interval for $-0.1876<E_j<-0.1707$, which is located 
beyond $E_j(L_{4,5})$.

%

\begin{figure}
\begin{center}
\resizebox{80mm}{!}{\includegraphics[angle=0]{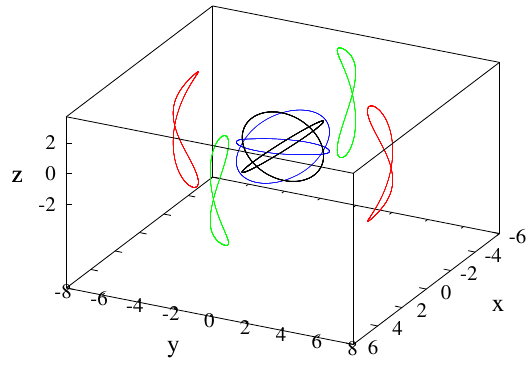}}
\end{center}
\caption{The pair of vLPOs around $L_{1,2}$ and the 
two periodic orbits of the vertical families around $L_{4,5}$ are given with 
red and green color respectively, at $E_j=-0.165$.
The two members of sao  and its symmetric 
family, as well as the two members of uao and its 
symmetric family are plotted with black and blue 
color respectively, for the same energy.} 
\label{pofig} 
\end{figure}

The morphology of the members of the vLPO family have a ``bent-eight'' 
morphology (red orbits in Fig.~\ref{pofig}), similar to the morphology of the 
vertical families around the stable Lagrangian points $L_{4,5}$ (green orbits in 
Fig.~\ref{pofig}). This is in agreement with the findings of \citet{op98} and 
\citet{rm09}. The depicted in Fig.~\ref{pofig} vLPO's, at $E_j =-0.165$, reach 
heights $|z|\approx 3.7$ away from the equatorial plane. In Fig.~\ref{pofig} 
we give also the representatives of two families introduced by \citet{hms82}, 
namely of the stable and unstable anomalous orbits (sao and uao respectively). 
The role of these families  will be examined in section~\ref{zetc}.  The positions of the periodic orbits of these families in the Poincar\'e section $z=0$ with $p_z>0$ (for value of energy $E_j=-0.165$) are given in table \ref{tab:models1}.

\section{The Phase Space}
\label{sec:start}

 We apply the method of color and rotation, in order to visualize the phase space structure 
 in the vicinity of vLPOs around $L_{1,2}$. For this reason, we choose 
to study the phase space structure in the vicinity of vLPOs $L_{1,2}$ for 
a value of energy $E_j=-0.165$, which is larger than the energy of the 
equilibrium points  $L_{1,2}$. We do so, because we want 
to study not only the phase space geometry close to vLPO1 and vLPO2 
but all the possible phase space objects that are emanated from the equilibrium points $L_3$ and $L_{4,5}$ and could affect the phase space structure close to vPLOs  around $L_{1,2}$. Thus, we need a value of energy at which representatives of all involved families of periodic orbits exist.

In our 4D space of section $z$=0, with $p_{z} > 0$ we input the four initial conditions
$(x,y,p_x,p_y)$ and we plot each orbit in a 3D subspace of the 4D space of section, e.g. in the 
subspace $(x,y,p_x)$ and we color the consequents according to their fourth 
coordinate, i.e. according to their $p_y$ values.

The phase space objects we encountered belong to two categories. The phase space objects of the first category are obtained if we perturb the initial conditions of the  vLPO1,2. These objects  are:
\begin{enumerate}
 \item {\bf Ring like structures} that are described in subsection \ref{dc}.  
 A kind of these phase space objects are reached  by applying  a perturbation to all 
four initial conditions in the ranges: $-5\leq \Delta x \leq5.2$, $-3.9 \leq 
\Delta y \leq 4$, $-0.02 \leq\Delta p_x \leq 0.09$ and  $-0.2\leq \Delta p_y 
\leq 0.2$. These phase space objects are due to the phenomenon of stickiness in chaos (see the subsection \ref{sti}).
 \item {\bf Clouds} result for larger perturbations as  described in subsection \ref{clouds}.  
\end{enumerate}

The phase space objects of the second category are obtained if we perturb the initial conditions of the  vLPO4,5. They are tori and sticky tori (see subsection \ref{tori}). Finally, the role of the z-axis family and its bifurcations to the structure of the phase space is analyzed in section \ref{zetc}.


\subsection{Ring like structures, aka double crescents}
\label{dc}
\begin{figure}
\begin{center}
\resizebox{90mm}{!}{\includegraphics[angle=0]{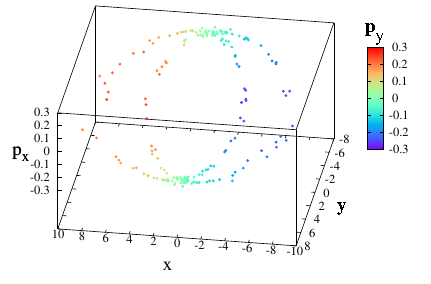}}
\end{center}
\caption{The 3D representation  $(x,y,p_x)$ of the double-crescent structure for 
the first 190 intersections with the space of section $z=0$ with $p_z>0$. Every point is 
colored according to the fourth dimension $p_y$ (color bar at the right-hand side of the figure).}
\label{thet1} 
\end{figure}

The first type of structure we encounter in the neighborhood of vLPOs around 
$L_{1,2}$ in phase space, is a ring-like one, resembling a   double crescent, 
with the two crescents facing and touching each other at their horns. Hereafter, 
we will briefly refer to this shape as the ``double-crescent''. The fourth 
dimension in this case follows the topology of this structure in the 3D subspace of the phase space as the color-rotation  method reveals. An example is 
given in Fig.~\ref{thet1}. In order to obtain the depicted orbit, we have 
considered the simple unstable vLPO at $E_j=-0.165$ and we applied a 
perturbation $\Delta y =10^{-4}$ in its initial conditions. However, we obtain 
similar double-crescent structures, by applying  perturbation: 
$-5\leq \Delta x \leq5.2$, $-3.9 \leq 
\Delta y \leq 4$, $-0.02 \leq\Delta p_x \leq 0.09$ and  $-0.2\leq \Delta p_y 
\leq 0.2$ 
For initial conditions outside these ranges we may find  clouds of points or we cross the 
surface of zero velocity and then we are beyond the phase space region 
in which we are allowed to integrate our orbits. (We remind that $E_j=-0.165> 
E_j(L_{4,5})> E_j(L_{1,2})$).

The structure depicted in Fig.~\ref{thet1} consists of the first 190 consequents 
of the orbit in the 4D space of section. These 190 consequents, correspond already to
$t=13$ Gyrs. For finding the chaoticity of these orbits we have to extend our calculations to unrealistic large time intervals. However, this allows us to understand the underlying dynamical mechanism and also helps us estimating the expected orbital dynamics in specific volumes of the phase space. It can be considered as confined within 
a narrow cylinder with dimensions about $[-10,10] \times [-8,8]$ in the $(x,y)$ 
plane  and a much smaller thickness of about 0.6 in the third dimension $p_x$.  
The $p_y$ ``colours'', evidently, follow a smooth color variation (cf. with the color bar 
at the right-hand side of the figure) which reflects the fact that the 
consequents are on, or at least close to, a smooth surface in the 4D space.
However, if we continue integrating the orbit, the consequents start diffusing 
in the surrounding phase space and we observe a mixing of colors. For other 
double crescent orbits, with initial conditions in the previously indicated ranges, 
this happens even earlier, when the orbits have formed only part of the 
double-crescent structure. At any rate, this evolution clearly indicates 
stickiness, or a weakly chaotic behaviour.

\begin{figure}
\begin{center}
\resizebox{80mm}{!}{\includegraphics[angle=0]{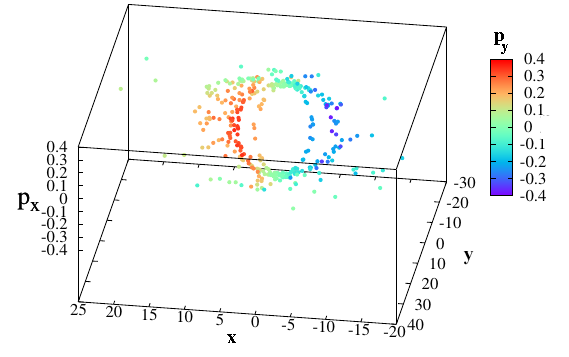}}
\end{center}
\caption{The 3D representation  $(x,y,p_x)$ of the double-crescent structure for 
382 intersections with the space of section $z=0$ with $p_z>0$. Every point is 
colored according to the fourth dimension $p_y$.}
\label{thet2} 
\end{figure}

When diffusion starts, the consequents form eventually a cloud where colors are 
mixed. The scattering of points in the 3D subspace and the mixing of colors 
indicate the strong chaotic behavior and the transition from the state of 
stickiness to the state of strong chaoticity. Furthermore, we observe  that the 
diffusion of points leads to the occupation of a larger volume in phase space. 
As we see in Fig.~\ref{thet2}, already for 382 intersections of the orbit with 
the space of section the occupied volume is 
$[x1,x2]\times[y1,y2]\times[p_{x1},p_{x2}]\times 
[p_{y1},p_{y2}]=[-20,25]\times[-30,40]\times[-0.4,0.4]\times[-0.4,0.4]$. This is 
a larger volume than the one of the narrow cylinder within which the first 190 
consequents are confined. We also observe that the 382 consequents drift 
away from the vLPOs mainly on the $(x,y)$ plane, while their $|p_y|$ values 
increase their extent by 33\% with respect to the values during the 
first 190 intersections. Eventually the points form clouds like these 
that are described in the next subsection.

\begin{figure*}
\begin{center}
\begin{tabular}{cc}
\resizebox{160mm}{!}{\includegraphics[angle=0]{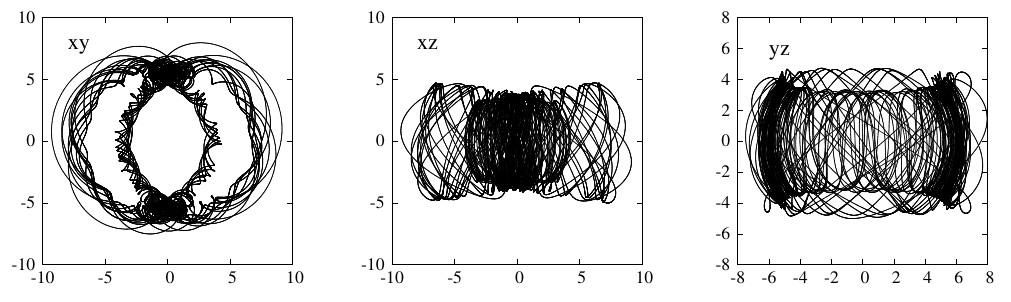}}
\end{tabular}
\caption{The $(x,y)$, $(x,z)$ and $(y,z)$ projections of the 
ring- or double-crescent-like orbit with initial conditions close to $L_1$ 
with $E_j=-0.165$, integrated for one Hubble time.}
\label{thetco}
\end{center}
\end{figure*}

The orbit in Fig.~\ref{thet1} and Fig.~\ref{thet2}, integrated over one Hubble 
time  in the $(x,y,z)$ space is given in 
Fig.~\ref{thetco}. The $(x,y)$ projection retains a double-crescent morphology 
similar to the one we have found existing in the $(x,y,p_x)$ space in Fig.~\ref{thet1}. 
Nevertheless, we have to keep in mind that in  Fig.~\ref{thet1} we depict a space of section,
while in Fig.~\ref{thetco} we plot the orbit in the configuration space.
It is remarkable that this overall morphology is one frequently observed in barred galaxies 
such as NGC~1326 and NGC~3504 ‎\cite{sb94}, or NGC~2665 (Red DSS image as in NED/NASA). 
Although this structure is expected to be supported by planar orbits 
\citep[see e.g. figure 21 in][]{Kaufmann1996self}, we see that the projection of sticky orbits reaching high $z$ retains this morphology.

As we can observe in the left panel of Fig.~\ref{thetco}, the orbit spends more 
time in the neighborhood of $L_1$ and $L_2$, which  results in the formation of 
two dense regions around the two unstable equilibrium points in the face-on, 
$(x,y)$, view. In the end-on, $(x,z)$ projection, these two dense regions 
correspond to the central, dense part, filling roughly a $(-4,4)\times(-4,4)$ 
region, while the orbit extends in the $x$-direction close to a distance from the center 
of about 8. Finally, in the side-on, $(y,z)$, projection, these two dense regions coincide 
with the two parenthesis-like arcs at the sides of this projection of the orbit 
(Fig.~\ref{thetco} right panel).


\begin{figure}
\begin{center}
\resizebox{80mm}{!}{\includegraphics[angle=0]{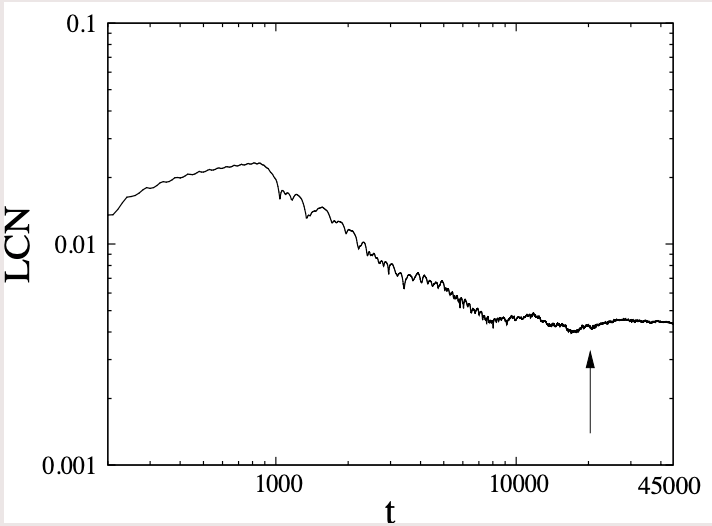}}
\end{center}
\caption{The variation of $LCN(t)$ (in log-scale) for the double-crescent 
structure. The arrow indicates the time at which  the consequents leave the 
double-crescent structure.}
\label{thetlcn} 
\end{figure}

In order to estimate the chaoticity of the orbit, we used the standard ``finite 
time'' Lyapunov Characteristic Number indicator \cite[see e.g.][section 2.10 and 
references therein]{gcobook}. For this we computed the quantity 
$LCN(t)=\displaystyle \frac{1}{t}\ln\left|\frac{\xi(t)}{\xi(t_0)}\right|$, in 
which $\xi(t_0)$ and $\xi(t)$ are the distances between two points of two nearby 
orbits at times $t = 0$ and $t$ respectively. The evolution of $LCN(t)$ with 
time is given in Fig.~\ref{thetlcn}. We observe that it has some fluctuations 
until the time at which the consequents leave the double-crescent structure 
(indicated with an arrow). Thereinafter, it increases slowly but constantly, 
tending to a positive value equal to $4.6\times 10^{-3}$.  This evolution 
confirms the sticky-chaotic character of the orbit.

\subsection{Clouds}
\label{clouds}
\begin{figure}
\begin{center}
\resizebox{80mm}{!}{\includegraphics[angle=0]{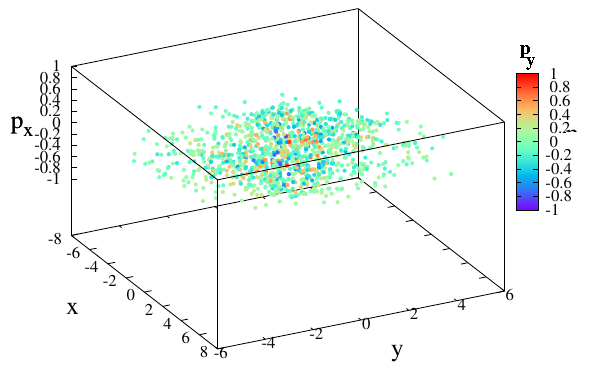}}
\end{center}
\caption{The cloud of points in the 4D space of section $z=0$ with $p_z>0$.
The consequents of the cloud are plotted in the 3D subspace $(x,y,p_x)$ and it 
is colored  according to the fourth dimension $p_y$. In the cloud, we observe mixing of 
colours.}
\label{cloudd} 
\end{figure}

\begin{figure}
\begin{center}
\resizebox{80mm}{!}{\includegraphics[angle=0]{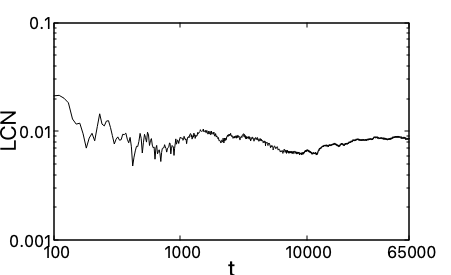}}
\end{center}
\caption{The variation of $LCN(t)$ for the orbit in Fig. \ref{cloudd} (in 
log-scale).}
\label{clouddlcn} 
\end{figure}

As we mentioned earlier, by increasing the absolute value of the perturbations, 
we reach initial conditions of orbits in the neighborhood of vLPOs, which 
rapidly demonstrate a chaotic character. Such orbits are represented in the 4D 
space of section by clouds of consequents with an irregular distribution in the 
3D subspaces. They also have an irregular distribution in the fourth dimension, 
reflected in the mixing of their colors. An example with a $\Delta y=-4$ 
perturbation, always for $E_j=-0.165$ is given in Fig.~\ref{cloudd}. This 
picture points to a strong chaotic behavior. The evolution of $LCN(t)$, given in 
 Fig.~\ref{clouddlcn}, confirms this conclusion, since, after 
some initial fluctuations, it tends to a relatively large positive number, i.e. to 
$10^{-2}$.


\begin{figure*}
\begin{center}
\resizebox{160mm}{!}{\includegraphics[angle=0]{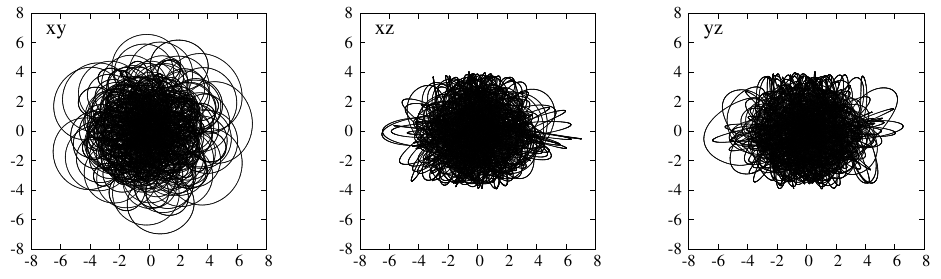}}
\caption{The $(x,y)$, $(x,z)$ and $(y,z)$ projections of a ``cloud orbit''
 in the configuration space (for one Hubble time).}
\label{clouddco}
\end{center}
\end{figure*}

The projections of this orbit in the configuration space can be described as 
roundish, with irregular shapes, as we can observe in  Fig.~\ref{clouddco}. The 
orbit is integrated again for a Hubble time, which corresponds now to 343 
intersections with the $z=0$ plane. These initial conditions lead to an 
irregular morphology of an orbit. Nevertheless, the orbit during the integration 
period, remains inside corotation. This is a counter-intuitive result, 
since the energy of the orbit is larger than the energy of the Lagrangian points 
at corotation, starting however well inside the corotation radius 
$R_c=R_{L_1}=6.31$.

Despite the fact that the  clouds of points indicate a strong chaotic behavior, 
their consequents remain confined  in the  4D space of 
section within a region $ [-8,8]\times[-6,6]\times[-1,1]\times[-1,1] $, which 
corresponds to a smaller volume than the region occupied by the 190 consequents 
of the sticky, double-crescent orbit in Fig.~\ref{thet1}. The origin and the 
conditions for this weak diffusion will be investigated below.

\subsection{Tori and sticky tori}
\label{tori}
\begin{figure}
\begin{center}
\resizebox{80mm}{!}{\includegraphics[angle=0]{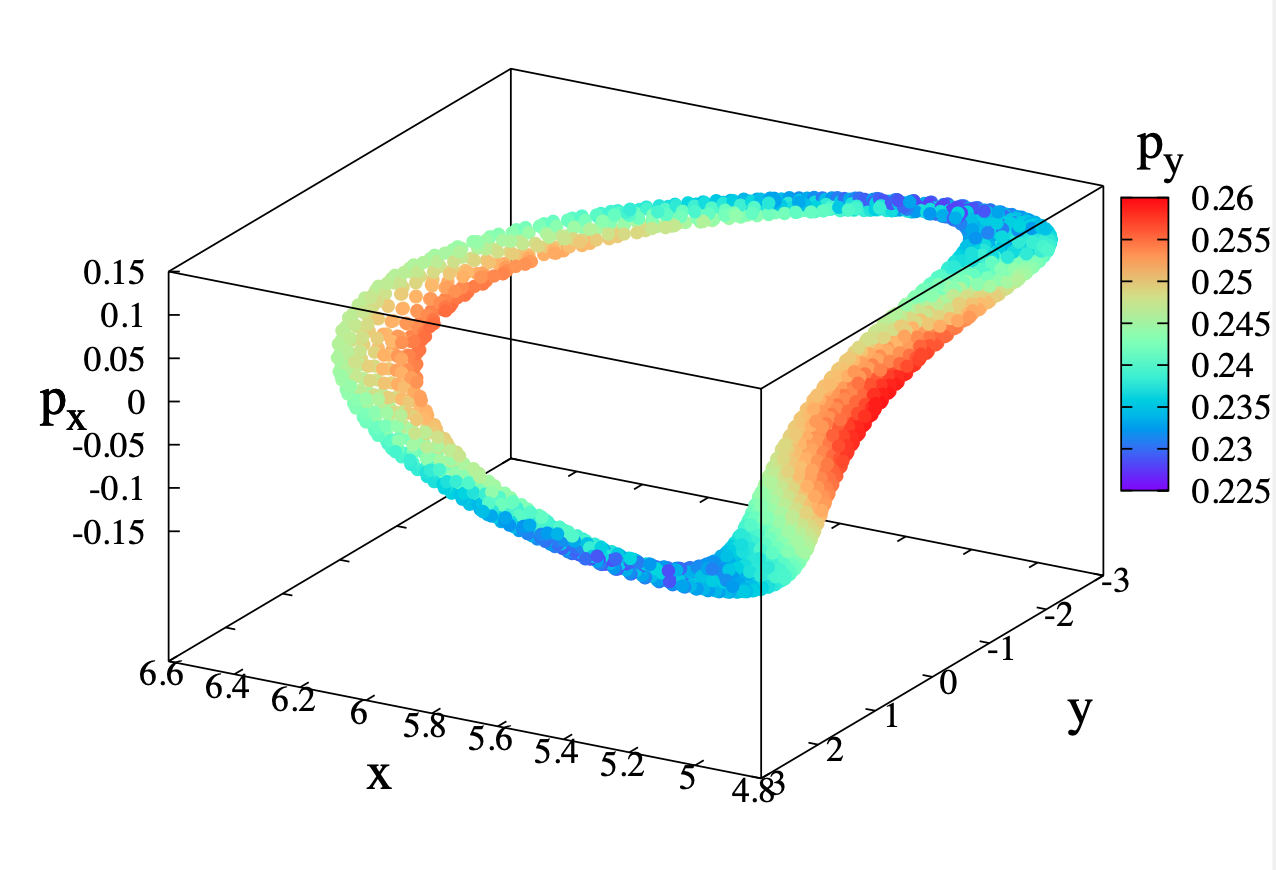}}
\end{center}
\caption{An invariant torus around the Lagrangian point $L_4$ in the 3D subspace
$(x,y,p_x)$ of the 4D space of section  $z=0$ with $p_z>0$, colored according 
to the $p_y$ values of the consequents. }
\label{tor} 
\end{figure}

According to the KAM theorem \citep{k54,a63, m62} in a near-integrable, 3D 
system, such as the one we study here, there are orbits that lie on 3D invariant 
tori close to stable POs. These invariant tori are 2D objects in the 4D space of section.
Such  objects are the invariant tori around the periodic orbits vLPO4,5 . The POs of these  families are stable and have initial conditions  $(x>0,0,0,p_y>0)$ 
(for $L_4$) and $(x<0,0,0,p_y<0)$ (for $L_5$). An example of these tori is given  in Fig.~\ref{tor}, depicting the perturbed by $\Delta x=0.6$ vertical PO around $L_4$.  This torus  has in the 4D space of section the morphology of a rotational torus \citep[][]{vetal97,vetal96}. This kind of tori have a smooth color variation on their 
surfaces, if we colour the consequents of the orbit according to their values in 
the fourth dimension \citep{kp11}.


\begin{figure}
\begin{center}
\resizebox{80mm}{!}{\includegraphics[angle=0]{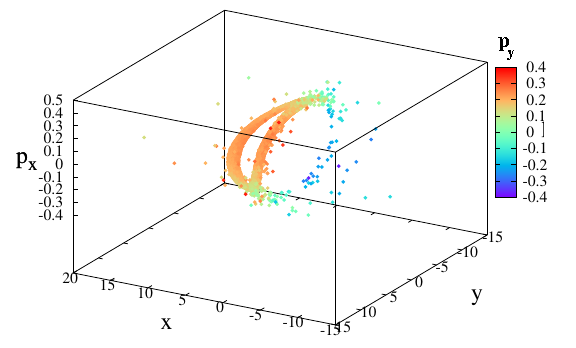}}
\end{center}
\caption{A sticky torus around the Lagrangian point $L_4$ in the 3D subspace  
$(x,y,p_x)$ of the 4D space of section  $z=0$ with $p_z>0$. The color 
represents the fourth dimension $p_y$. The consequents depart from the sticky 
torus after 17120 intersections.}
\label{ttor} 
\end{figure}

By increasing further the perturbation of the initial conditions of the vLPO4 
 along the $x$-direction, we encounter a different dynamical behaviour for 
$\Delta x=1.3$. In this case we initially find  again a rotational-torus-like 
structure considering the first 17120 consequents, but for larger integration 
times the consequents  start occupying a larger volume around the periodic orbit
 (Fig.~\ref{ttor}). This is the phenomenon of stickiness on  tori (see introduction), 
 as encountered in the case of the tori around stable POs in 
  3D autonomous Hamiltonian systems.  During the sticky period of an orbit it forms 
  a torus-like orbit, which was called sticky torus in \citet{kp11}.  
  Such tori are unstable and present a weak 
chaotic behavior. For usual time scales in Galactic Dynamics, the dynamical behavior of 
such orbits can be considered practically identical to that of regular orbits.



%

\begin{figure}
\begin{center}
\resizebox{80mm}{!}{\includegraphics[angle=0]{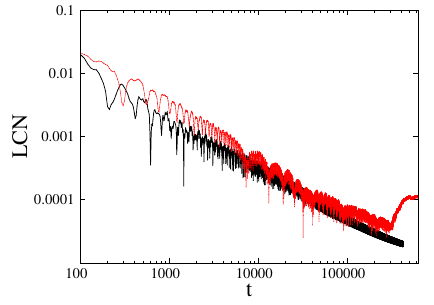}}
\end{center}
\caption{The variation of LCN(t) (with black color) for the case of the 
invariant torus which is depicted at the Fig. \ref{tor}. The variation of 
$LCN(t)$ (with red color) for the case of the sticky torus which is depicted at 
the Fig. \ref{ttor}. Both axes are in log-scale.}
\label{ttorlcn} 
\end{figure}

This weak chaotic behavior can be also identified by means of the finite time 
$LCN(t)$. However, we need an even larger integration time, than the integration 
time used when applying the method of color and rotation, in order to appreciate 
it. In Fig.~\ref{ttorlcn}, we observe that the $LCN(t)$, which corresponds to 
the invariant torus of the Fig.~\ref{tor} (black curve) tends to zero. During the 
same time interval, the $LCN(t)$ that corresponds to the sticky torus of 
Fig.~\ref{ttor} (red curve), initially decreases, behaving like the $LCN(t)$ of 
an ordered orbit. However,  at t=315000, it starts increasing and finally it tends to 
a relatively small positive number $10^{-4}$. The $LCN(t)$ identifies the weak 
chaotic behavior of the sticky torus after 325000 time units, thus this index traces 
the chaotic behaviour slower than the method of color and rotation. 

The difference between this kind of sticky behaviour and the one of the orbit 
described in section~\ref{dc} is explained in the following paragraph. 


\subsection{Manifolds around vLPOs and  stickiness in chaos}  
\label{sti}

\begin{figure*}
\begin{center}
\resizebox{160mm}{!}{\includegraphics[angle=0]{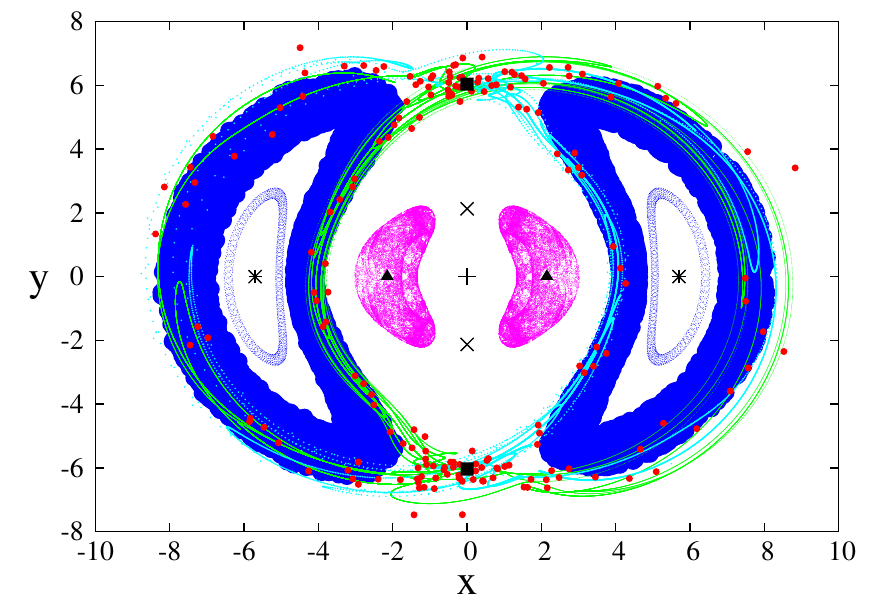}}
\end{center}
\caption{The $(x,y)$ projection of the $(x,y,p_x,p_y)$ 4D space of section, 
for consequents and invariant manifolds associated with the diffusion of 3D 
orbits form the neighborhood of vLPOs around $L_{1}$ and $L_{2}$ for 
$E_J=-0.165$. We consider $z=0$ with $p_z>0$. The initial conditions of the two 
vLPOs are indicated with black squares, while those of the vertical POs around 
$L_{4,5}$ with black stars. Furthermore, the central cross corresponds to the 
initial conditions of the POs of the z-axis family (coincides with the location 
of $L_{3}$), black triangles mark the location of the two sao POs and the 
$\times$ symbols those of the two uao ones. The 190 consequents of the 
double-crescent orbit are given with heavy red dots. The 2D projection of the 
unstable invariant manifolds of the simple unstable vLPOs around $L_{1}$ and 
$L_{2}$ are drawn with green and cyan color respectively. The projections of the 
tori around the stable $L_{4,5}$ are drawn with blue, while those around the sao 
POs with magenta color. This landscape offers an illustrative representation of 
the main structures existing in 
the phase space we study.}
\label{all2d} 
\end{figure*}

An illustrative way for understanding the structures we encounter in phase 
space, due to the presence of the vertical families in the neighborhood of $L_1$ 
and $L_2$ for $E_j=-0.165$, is offered by the $(x,y)$ projection of the 4D phase 
space $(x,y,p_x,p_y)$. This is presented in Fig.~\ref{all2d}. The initial 
conditions of the vLPOs around $L_1$ and $L_2$ are marked with two black squares at 
$y>0$ and $y<0$ respectively.  Besides the vertical families around $L_{1,2}$ we have also the 
vertical families around the stable equilibrium points $L_{4,5}$. In Fig.~\ref{all2d} their 
locations are indicated with two black ``\raisebox{-.5ex}{*}'' symbols. 

In Fig.~\ref{all2d}, apart from  the location of the POs and the projections of the 
tori, we also plot the  projections on the $(x,y)$ plane 
of the asymptotic 
curves of the unstable invariant manifolds of the two simple unstable vLPOs 
around $L_{1,2}$. We plot with green the vLPO around $L_1$ and with cyan the 
vLPO around $L_2$. The numerical construction of the asymptotic curves is 
described in detailed in \cite{kpc13}. The unstable invariant manifolds of 
the simple unstable periodic orbits are 1-dimensional objects in the 4D space of 
section,  because they correspond to one eigenvalue, which is outside the unit 
circle \citep[][section 34A]{arn1}. Due to the presence 
of higher-dimensional objects, namely due to the (blue) invariant tori around 
the stable vLPO4 and vLPO5, the asymptotic curves are wrapped 
around them in the 2D projection of the space of section \citep[a similar case 
has been encountered in][]{kpc13}. 
By rotating the figure in 3D subspaces, e.g. 
in ($x,y,p_x$), we 
realize that the manifolds warp around, without intersecting, the blue tori at a 
given $E_j$. The first 190 points of the double-crescent structure (red dots in 
Fig.~\ref{all2d}), stick on the asymptotic curves of the invariant manifolds in 
the 3D, and consequently on the 2D, projections of the 4D space of section. As 
we can see in Fig.~\ref{thet1}, these 190 consequents have also a smooth color 
variation, reflecting a smooth distribution of their fourth coordinate $(p_y)$. 
This sticky-on-manifolds behaviour of the 190 consequents, shows once again,
as in \cite{kpc13}, the presence of the  phenomenon of stickiness in 
chaos in galactic type Hamiltonian systems.

\subsection{The role of z-axis and its bifurcations}
\label{zetc}
Finally,  for completing our study,  we investigate the important for our study 
``z-axis'' orbits and its bifurcations. The POs of 
the z-axis family lie entirely along the rotating z-axis, i.e.  and can be considered as 
the vertical  Lyapunov family of the  Lagrangian point $L_3$. They have initial conditions 
$(x,y,p_x,p_y)=(0,0,0,0)$ in the 4D space of section. \citet{hms82} computed 
this family as well as its two bifurcations, which they named stable and 
unstable anomalous orbits (sao and uao respectively). Both of these families 
have at each energy two symmetric representatives. Specifically, the two sao 
orbits have  initial conditions $(x>0,0,0,p_y<0)$ and $(x<0,0,0,p_y>0)$ and 
the two uao $(0,y>0,p_x>0,0)$ and $(0,y<0,p_x<0,0)$. The dynamics and the stability 
of these 
families have been later studied  by  \citet{mp87} and by \citet{pz90}.

\begin{figure}
\begin{center}
\resizebox{70mm}{!}{\includegraphics[angle=0]{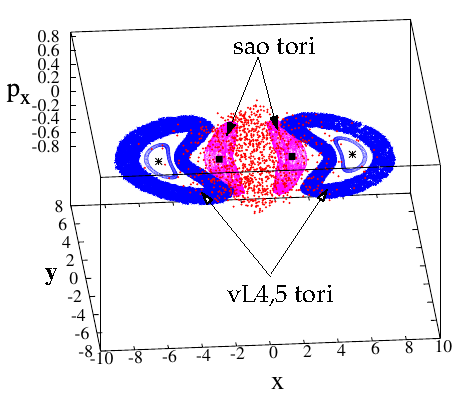}}
\end{center}
\caption{The 3D $(x,y,p_x)$ projection of the cloud (red dots) in 
the 4D space of section $z=0$ with $p_z>0$ for $\Delta y 
=10^{-3}$, always for $E_J=-0.165$ . The 3D projection of the invariant tori around the vertical POs around $L_{4,5}$ and  the 3D projection of the invariant tori around the POs of the two sao families are represented by blue and magenta color 
respectively.}
\label{cloud3d} 
\end{figure}

The z-axis family is for large regions of the parameter space unstable and in 
fact complex unstable \citep[][]{mp87,pz90}. This is the case also for its  
representative at $E_j=-0.165$. We find that even a tiny perturbation of its 
initial conditions leads to chaotic orbits such as the one in Fig.~\ref{cloudd}. Here we underline the fact that the cloud in  Fig.~\ref{cloudd} can be obtained by applying a perturbation $\Delta y =-4$ at the initial condition of vLPO1 or applying a $\Delta y =2.035$ perturbation  at the initial condition of the z-axis family.
The z-axis orbits, perturbed along the $y$-direction, keeping the rest of the initial 
conditions equal to zero, demonstrate this behavior for $|y_0|<2.5$. However, 
these chaotic orbits do not easily cross the $L_{1,2}$ regions to visit the zone 
of the model beyond corotation. The origin of this behaviour becomes apparent  by 
inspection of Fig.~\ref{cloud3d}, where we present an orbit perturbed by 
$\Delta y = 10^{-3}$.  We remind that we observe the same  behaviour also for larger perturbations. For example we obtain the same orbit depicted  in Fig.~\ref{cloudd}, also by applying a  $\Delta y =2.035$ to the initial conditions of the z-axis orbit.
The existence of the magenta invariant tori around 
the two sao orbits, assisted by the presence of the blue invariant tori around 
the vertical POs around $L_{4,5}$ hinder the diffusion of the chaotic orbits and 
keep them trapped in the central region of the $(x,y,p_x)$ projection.  The few consequents that overcome the two tori  barriers stick on the blue tori that are around the v$L_{4,5}$ (Fig. \ref{cloud3d}).

\begin{figure}
\begin{center}
\resizebox{90mm}{!}{\includegraphics[angle=0]{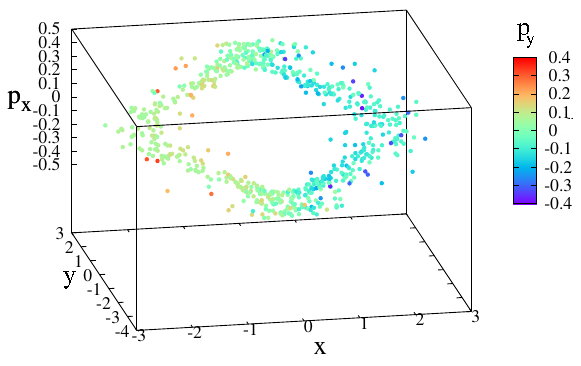}}
\end{center}
\caption{A 4D representation of the space of section for the first 700 
consequents of the toroidal zone. The spatial coordinates are $(x,y,p_x)$, 
while the colour of the consequents represents their $p_y$ value.}
\label{zone1} 
\end{figure}

\begin{figure*}
\begin{center}
\resizebox{160mm}{!}{\includegraphics[angle=0]{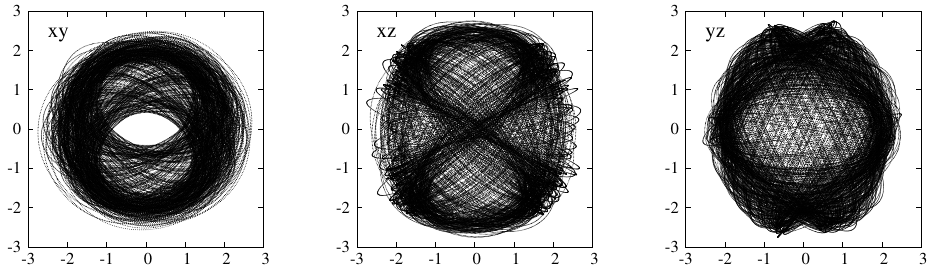}}
\caption{The $(x,y)$, $(x,z)$ and $(y,z)$ projections of the orbit 
corresponding to the ``toroidal zone'' of Fig.~\ref{zone1}.
integrated for one Hubble time.}
\label{zoneco}
\end{center}
\end{figure*}

Perturbations of the bifurcations of the z-axis (sao and uao) lead to different structures in 
phase space, as expected, since the sao orbits are stable while those of the uao 
families are unstable. Perturbations of sao lead to quasi periodic orbits, the 
tori of which have been presented in Fig.~\ref{all2d}. On the other hand,  by 
adding even tiny perturbations, like $\Delta x=10^{-3}$, in the initial 
conditions of the uao POs, we find in the 3D $(x,y,p_x)$ subspace ring-like 
objects. Such an orbit, with the aforementioned perturbation at $E_j=-0.165$, is 
given in Fig.~\ref{zone1} and we will refer to it as  a ``toroidal structure''. 
Despite the fact that, vaguely speaking, it has a toroidal topology, 
its actual  shape is irregular. The consequents do not lie  on a smooth surface in this 3D projection.We observe that the consequents building this structure 
have a rather smooth color variation, albeit some points already are departing from it 
at the end of the integration time. All these point to a weakly chaotic orbit. 
Its morphology in the configuration space, integrated over a Hubble time, is 
given in Fig. \ref{zoneco}. It remains confined in a $(x,y,z)$ region 
$[-3,3]\times[-3,3]\times[-3,3]$, with a toroidal face-on and a spheroidal 
edge-on view.


\begin{figure}
\begin{center}
 \resizebox{85mm}{!}{\includegraphics{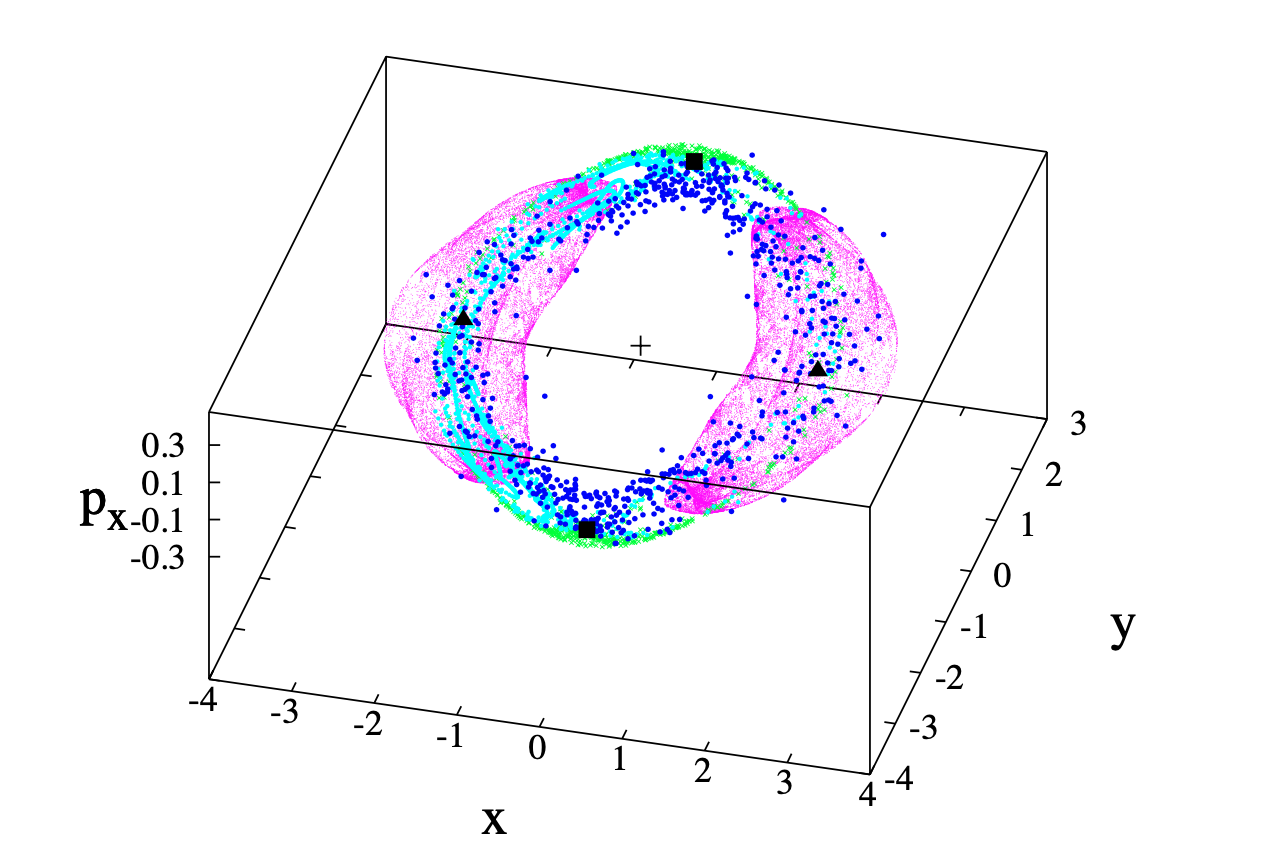}}
\caption{The 3D, $(x,y,p_x)$, projection of the unstable invariant 
manifolds 
of the simple unstable POs of uao and its symmetric family for $E_j=-0.165$, 
are plotted with 
green and cyan color respectively. The tori around the stable POs of sao and 
its 
symmetric family are drawn with magenta color. The blue points are the 
consequents of the toroidal zone of Fig.~\ref{zone1}, as projected again in the 
3D, $(x,y,p_x)$ subspace of the space of section. The locations of  the POs of the z-axis family, 
of the two uao's and of the two sao's are indicated with a cross, two black 
squares and two black triangles respectively.}
\label{zone3d}
\end{center}
\end{figure}

We computed the unstable invariant manifolds of the simple unstable uao POs,
as well as of its symmetric family, and we present them for $E_j=-0.165$ in 
 Fig.~\ref{zone3d}, with green and cyan color 
respectively. For the computation we followed again the 
steps described in detail 
in \citet{kpc13}. Also in this case the unstable asymptotic curves of a simple 
unstable PO are guided from the invariant tori that exist in the region. In the 
present case, they are guided by the invariant tori around the two sao orbits, 
drawn with magenta color in Fig. \ref{zone3d}.

The first 700 consequents of the toroidal structure depicted in Fig.~\ref{zone1} 
stick on the unstable invariant manifolds of the two uao POs (Fig. 
\ref{zone3d}). We observe again the presence of stickiness in chaos, as in the 
case of the double-crescent structure (Fig. \ref{all2d}). However, as expected 
for a sticky orbit, by continuing the integration, we have a transition from the 
toroidal-like structure depicted in Fig. \ref{zone1}, to a cloud of points, with mixing of colors 
in the fourth dimension and a strongly chaotic behaviour of the orbit.


\begin{figure}
\begin{center}
\resizebox{80mm}{!}{\includegraphics[angle=0]{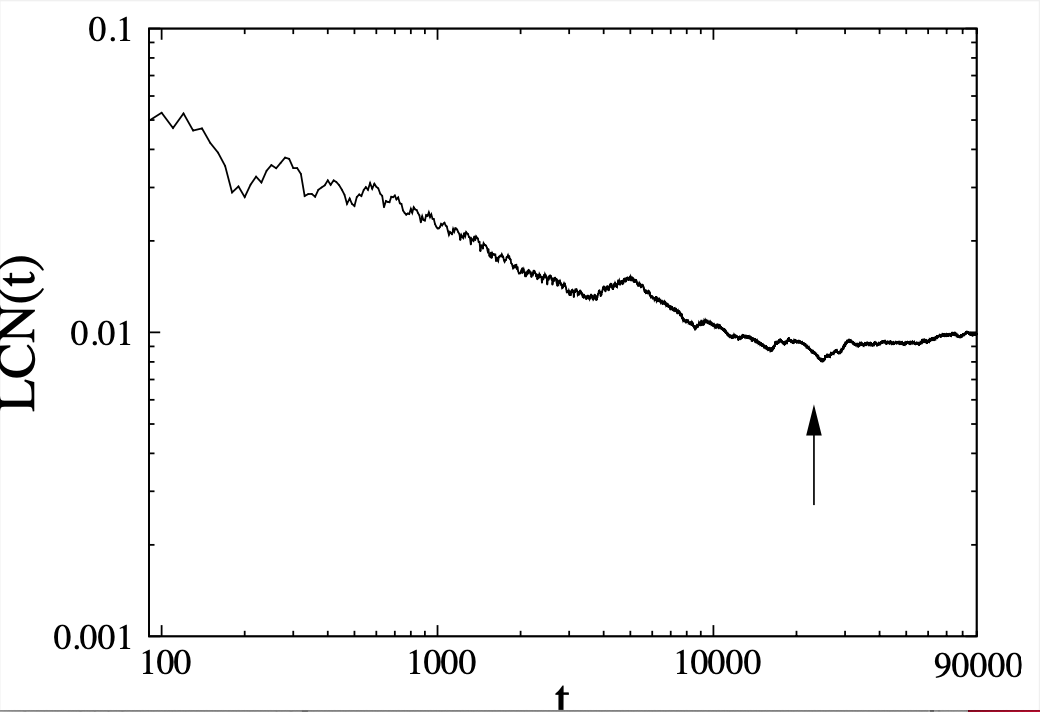}}
\end{center}
\caption{The variation of $LCN(t)$ (in log-scale) for the orbit in 
Fig.~\ref{zoneco}. The arrow indicates the time $t=25000$ at which the points 
depart from the toroidal structure in Fig.~\ref{zone1}.}
\label{zonelcn} 
\end{figure}

This transition is reflected also in the variation of its $LCN(t)$. In 
Fig.~\ref{zonelcn}  we observe a behaviour of this chaos indicator similar 
to that of the orbit in Fig.~\ref{thetlcn}. Beyond the point indicated with 
an arrow, it levels off at a positive number of the order of $10^{-2}$. 
Nevertheless, again due to the presence of the magenta tori in Fig.~\ref{zone3d}, 
the consequents of the toroidal structure remain confined  for a long time in a restricted zone of 
phase space and this is also compatible with the morphology of the orbit in the 
configuration space (Fig.~\ref{zoneco}).

\section{Diffusion}
\label{sec:start1}

We come now to the study of the diffusion of chaotic orbits located in the 
neighborhood of the vLPOs around $L_{1,2}$. The four types of chaotic orbits we 
discussed in Section~\ref{sec:start} are the double-crescent structures, the 
clouds of points, the sticky tori (that are around vLPO4,5 and sao orbits)
 and the toroidal structures close to the periodic orbits of uao and its symmetric one. 
 For these types of 
orbits we computed their diffusion speed as defined in \citet{kpc13}. The estimation of 
the diffusion is based on the estimation of $V_d$, the diffusion volume. For 
this we compute, as time increases, the volume of the  hypersphere on the 4D 
space of section, with center  the position of the periodic orbit and  radius 
the mean distance of the consequents from the periodic orbit. The diagram of  
$V_d$ versus $t$ gives us valuable information about the diffusion of 
the orbits. We call $T_d$ (diffusion time) the time needed by an orbit to reach the maximum 
$V_d$ ($V_{max}$), while the ratio $V_{max}/T_d$ gives us what we call ``diffusion speed'',  
$u_d$ \citep{kpc13}. We underline the fact that the most important information is 
obtained through the diagram $V_d$ versus $t$ (the information given  by $V_{max}$ and by the diffusion speed 
is complementary). This diagram can distinguish different kinds of orbital behavior (regular, weakly chaotic, and strong chaotic) giving in parallel the time evolution of the diffusion  \citep[see][]{kpc13}. The maximum diffusion volume and 
the diffusion speed give us only an indication of the maximum diffusion of the orbits and they have a complementary role.   

\begin{figure*}
\begin{center}
\begin{tabular}{cc}
\hspace{-15mm}
a)
\hspace{3mm}
\resizebox{200mm}{!}{\includegraphics{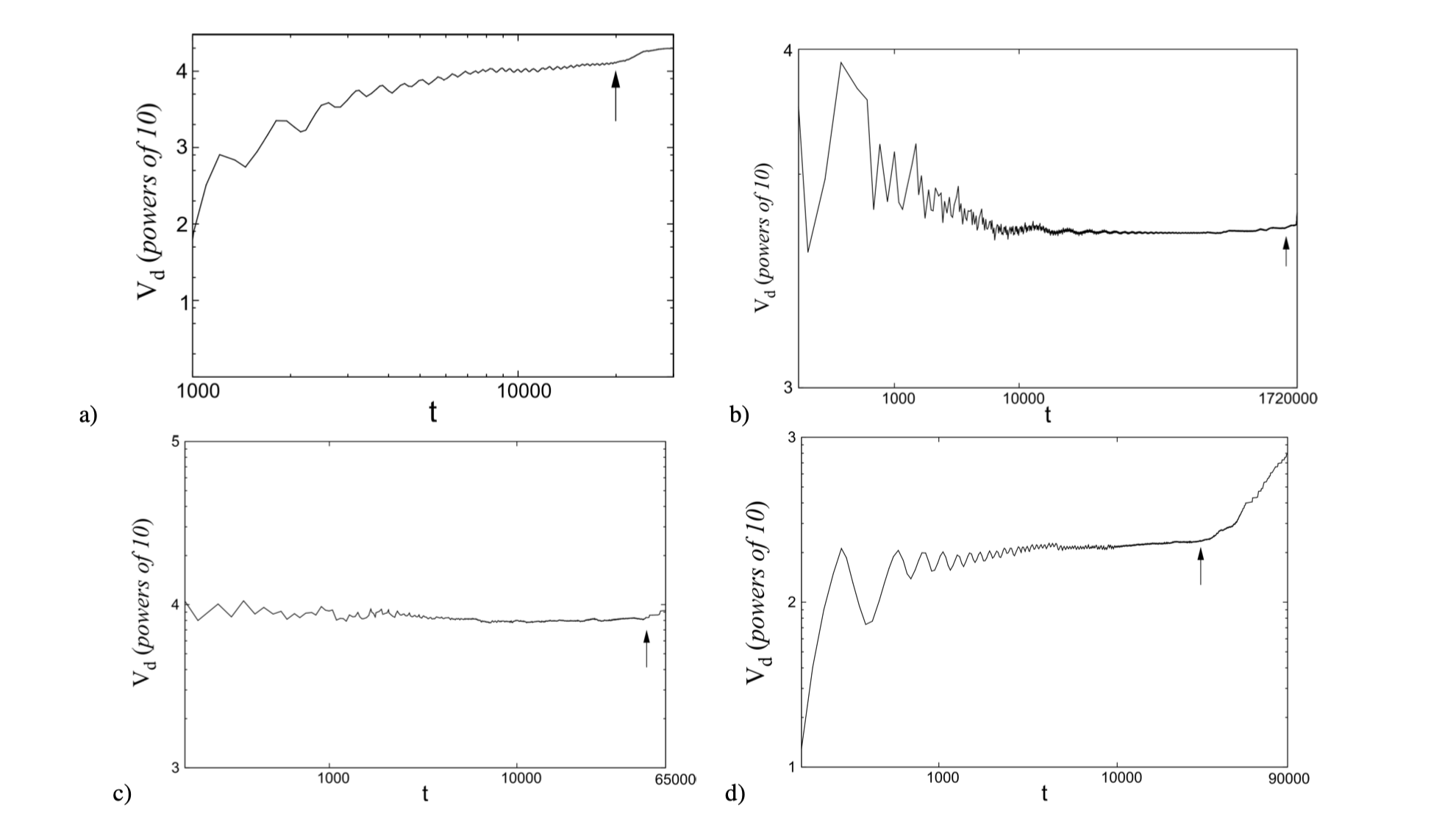}}\\
\end{tabular}
\caption{\textbf{a)} The time evolution of $V_d$ (in log-scale) in the case of 
the double-crescent structure for $1000 \leq t \leq 30000$. \textbf{b)} The time evolution of $V_d$ (in log-scale) at the case of  the sticky torus for $10 \leq t \leq 1720000$. \textbf{c)} The time evolution of $V_d$ (in log-scale) at the case of the cloud for $10 \leq t \leq 65000$. \textbf{d)} The time evolution of $V_d$ (in log-scale) at the case of 
the toroidal zone  for $10 \leq t \leq 90000$.} 
\label{diftheta}
\end{center}
\end{figure*}

Based on the above definitions, we first compute the diffusion speed of the orbit 
building the  double-crescent structure 
(Fig.~\ref{thet1}). In Fig.~\ref{diftheta}a we give the variation of $V_d$ with 
time. At the very beginning of the integration, the diffusion volume  increases 
almost monotonically, then it continues increasing fluctuating tending to a 
value $V_d \approx 13000$, up to the point indicated with an arrow. Until that 
time, the orbit remains sticky to the unstable invariant manifolds of the vLPOs 
around $L_{1,2}$, which are guided by the invariant tori around $L_{4,5}$. For 

larger time,  the consequents depart from the double-crescent structure 
(corresponding to 190 intersections with the space of section).  We observe a 
characteristic step in the variation of $V_d$ at this point (Fig.~\ref{diftheta}a).   From this point 
on, the diffusion  volume increases further, tending to a value 
$V_{max}\approx28100$, which is reached for $T_d=45000$. Thus, the diffusion 
speed of the double-crescent structure  is 0.624. The consequents will occupy 
larger volumes in the phase space as time increases
because the system is open (the value of energy is above the value of energy $E(L_{1,2}$) 
and the consequents cannot be restricted  from the curve zero velocity surface). 
This results the increasing  of diffusion volume versus time.


The time evolution of the diffusion volume of the sticky-torus-orbit around 
$L_4$ (Fig.~\ref{ttor}) is given in Fig.~\ref{diftheta}b. The diffusion volume is 
constant for a  large time interval ($10000 < t < 45000$). Once
the consequents depart from the torus region, at the time  
indicated with an arrow at the right-hand side of Fig.~\ref{diftheta}b, we 
observe a small increase of $V_d$. This small increase will continue forever because 
the system is open (the value of energy is above the value of the $E(L_{1,2}$) energy
and the consequents cannot be restricted  from the curve zero velocity surface). 
We compute the diffusion  speed for this orbit to be $u_d=0.001034$ 
($V_{max}=1765$ and $T_d=1707000$).




The orbit that gives the clouds in the space of section (Fig.\ref{cloudd}) is 
strongly chaotic, but, as long as we integrated it, it remains trapped in 
a subvolume of the phase space, due to the 
presence of the invariant tori of the sao's. In this case, the evolution of the 
diffusion volume is given in Fig.~\ref{diftheta}c. The values of the volume level off at about 
$V_d=8000$, but close to the end of the integration, beyond a time indicated 
with an arrow, there is a small further increase. This increase is due to the 
few points that slip through the ``holes'' between the tori, reaching larger 
distances. The consequents occupy a maximum diffusion volume $V_d=9200$ at 
$T_d=62000$, which gives a  diffusion speed $u_d=0.14832$.



 
Finally, the evolution of the diffusion volume for the orbit of the toroidal 
structure 
(Fig.~\ref{zone1}) is depicted in Fig. \ref{diftheta}d. The 
toroidal-structure-orbit is initially sticky to the unstable invariant 
manifolds of the two uao's, which are guided through the invariant 
tori of either 
sao's.  As we see in Fig.~\ref{diftheta}d $V_d$ increases with a decreasing 
rate up to the point that is indicated with an arrow, tending to level off. 
The indicated time corresponds again to 
the characteristic time at which the points of the toroidal structure leave the 
unstable manifolds, forming a cloud. We observe that, the diffusion volume 
increases, reaching a maximum value $V_d=790$ at $T_d=90000$, 
giving $u_d=0.0088$. In the diagram of Fig.~\ref{diftheta}d we observe that the increasing 
rate is larger than this of the cloud in Fig. \ref{diftheta}c. This is because the consequents are stuck initially in a smaller volume (the unstable invariant manifolds of the two uao's, which are guided through the invariant tori of either sao's) than the one occupied by the cloud in the previous case. When they expand outside this volume, they start to occupy larger volumes in  phase space and the increasing diffusion rate is larger than this of the clouds.  This is due to the fact that initially the volume of these consequents was smaller.




\section{DISCUSSION AND CONCLUSIONS}
\label{sec:conclusions}

The present study underlines the significance of the existing structure of 
the phase space in the neighborhood of POs for the 
determination of the orbital behaviour of  orbits in the vicinity of the POs. Apart from the 
case, in which the perturbation of the initial conditions of a PO
brings the perturbed orbits on nearby tori of other families, there are several 
kinds of stickiness, that may keep the orbit confined in the vicinity of a PO 
for times, dynamically significant in Galactic Dynamics. This may delay the 
diffusion of the orbit in remote phase space regions, in the same way that the
trapping of a chaotic orbit in a restricted volume may even reinforce morphological 
features observed in disk galaxies \citep{pk14a, p22}. 
%
%
%
%
%

\vspace{0.5cm}

By perturbing the initial conditions of the vLPOs around the Lagrangian points 
$L_{1,2}$, we found five types of orbital behaviour:
\begin{enumerate}
\item  \textbf{Double-crescent, sticky chaotic structures:} The first type 
is 
represented by double-crescent structures with smooth color variation along 
their consequents in the 4D spaces of section. This means that they are at least close to a smooth surface in the 4D space. This morphology lasts for long integration times 
(of the order of a Hubble time). However, the points forming them are sticky on the unstable 
invariant 
manifolds of the simple (or double) unstable POs around the Lagrangian points 
$L_{1,2}$. The manifolds are guided from the invariant tori that are 
found around the 
stable vertical periodic orbits associated with the Lagrangian points 
$L_{4,5}$.  For larger integration times the points diffuse away of these structures and occupy 
larger volumes in the phase space. Thus, eventually, they are chaotic with large 
diffusion speeds.

 \item \textbf{Clouds:} The second orbital type can be described as 4D clouds of 
points in the 4D spaces of section. The consequents have irregular distribution 
in the 3D projections and mixing of colors when colored according to their 
values in the fourth dimension. This is typical of a strong chaotic behavior, as 
also reflected  in the large values of the Lyapunov characteristic numbers 
$(LCN)$ of these orbits. Nevertheless, their expansion rate is decelerated, by 
the presence of the invariant tori around the sao POs. This keeps their 
$u_d$ values relatively small.

\item \textbf{Quasi-periodic orbits on tori:} The third type are regular orbits 
forming tori with smooth color variation along their surfaces. In this case, 
their $LCN$ tends to zero. These are quasi-periodic orbits  around the stable 
vLPO4 and vLPO5 or around either branch of the sao's.
 
\item \textbf{Sticky tori:} The fourth type are the sticky tori. In the 4D 
spaces of section, these orbits stick on the tori around the stable vLPO4 and 
vLPO5 for long time. Eventually, these weakly 
chaotic orbits diffuse in phase space with small diffusion speeds.

\item \textbf{Toroidal zones/structures:} Finally,  we find the orbits we call toroidal 
zones or toroidal structures. These are orbits, sticky on the unstable invariant 
manifolds of the 
two simple unstable uao POs, which are guided by the  invariant tori around the sao's. 
Up to a certain time the consequents of these orbits stay on a quasi-regular 
toroidal-like zone retaining a smooth color variation along their surface. Then, 
this weakly chaotic behaviour turns to a strongly chaotic one, with totally 
irregular 
distribution of the consequents in the 3D projections, mixing of colors and 
large positive $LCN$. They are practically clouds of scattered points in the 4D 
spaces. Nevertheless, the diffusion of these orbits is small due to the 
presence of the  the sao tori in the region.


\end{enumerate}

Stickiness is ubiquitus in the neighborhood of the simple unstable 
vLPOs around $L_{1,2}$. Stickiness in chaos appears typically in orbits sticky 
to the asymptotic curves of the vLPO1,2 unstable invariant manifolds. They remain 
close to the tori of the vertical families of POs associated with $L_{4,5}$. 
Stickiness in chaos is also found in orbits remaining sticky to the asymptotic 
curves of the unstable invariant manifolds of the uao POs. In addition we also 
find sticky orbits to the sao, as well as to the vLPO4 and vLPO5, tori. In 
conclusion, the presence of the sao tori is very important for restricting the 
orbital diffusion from the region around the vLPOs.


A basic, general, conclusion of this study is that the presence of stickiness in 
the orbital behaviour of the orbits in our 3D Hamiltonian system, hinders the 
diffusion of chaotic orbits. We find that strongly chaotic orbits, with large 
$LCN$, have small diffusion speeds. Such orbits are the clouds and the orbits we 
called toroidal zones/structures. Contrarily, weakly chaotic orbits like those 
that build the double crescent structures have large diffusion speeds. The 
difference between the two cases is the existence (former case) or the absence 
(latter case) of barriers, i.e of nearby invariant tori, in the vicinity of the 
POs. Thus, although in the 6D phase space of a 3D system we can follow more 
paths, compared to a 2D system, for visiting all regions of a chaotic sea, in 
practice this is proven to be in many cases problematic. The complexity of the 
phase space plays a crucial role, since in the vicinity of the POs one 
can find several invariant structures (tori or manifolds) which in effect delay 
the diffusion of chaotic orbits.

The lack of observed structures  away of the equatorial plane at the corotation region in galactic images indicates the absence of families that can act as obstacles for the diffusion of the orbits in the surrounding chaotic sea. This refers mainly to the sao and uao families which therefore are probably not populated in real galaxies.

\vspace{0.5cm} 
\noindent \textit{Acknowledgements}

We thank Prof. G.~Contopoulos for fruitful discussions and very useful comments.

\vspace{0.5cm} 
\noindent \textit{Data Availability}

The data underlying this article will be shared on reasonable request to the corresponding author.




\label{lastpage}

\end{document}